\documentstyle[rotate,psfig]{lamuphys}
\makeatletter
\let\chapter\hid@chapter
\newcommand{\n}{\noindent}
\newcommand{\bc}{\begin{center}}
\newcommand{\ec}{\end{center}}
\newcommand{\bi}{\begin{itemize}}
\newcommand{\ei}{\end{itemize}}
\newcommand{\bd}{\begin{description}}
\newcommand{\ed}{\end{description}}
\def\ghz[#1]{$\nu = #1$ GHz}
\def\mhz[#1]{$\nu = #1$ MHz}
\def\r[#1,#2]{$r = (#1 \pm #2)$ mas}
\def\bet[#1,#2]{$\beta_{app} = (#1 \pm #2)$}
\def\m[#1,#2]{$\mu = (#1 \pm #2)$ mas/yr}
\def\p[#1,#2]{P.A. = $(#1 \pm #2)\,^{\circ}$} 
\def\etal{{\it et al.\,}}

\def\hub{$H_0=100$ km s$^{-1}$ Mpc$^{-1}$, $q_0 = 0.5$}

\def\sprop{$S_{\nu} \propto \nu^{\alpha}$}
\def\solar{\ifmmode_{\mathord\odot}\else$_{\mathord\odot}$\fi~}
\def\deg{\ifmmode $\setbox0=\hbox{$^{\circ}$}$^{\,\circ}
          \else    \setbox0=\hbox{$^{\circ}$}$^{\,\circ}$\fi\,}
\def\pdeg{\ifmmode $\setbox0=\hbox{$^{\circ}$}\rlap{\hskip.11\wd0 .}$^{\circ}
          \else \setbox0=\hbox{$^{\circ}$}\rlap{\hskip.11\wd0 .}$^{\circ}$\fi~}
\def\arcs{\ifmmode {^{\scriptscriptstyle\prime\prime}}
          \else $^{\scriptscriptstyle\prime\prime}$\fi~}
\def\arcm{\ifmmode {^{\scriptscriptstyle\prime}}
          \else $^{\scriptscriptstyle\prime}$\fi~}
\def\la{\mathrel{\mathchoice {\vcenter{\offinterlineskip\halign{\hfil
$\displaystyle##$\hfil\cr<\cr\sim\cr}}}
{\vcenter{\offinterlineskip\halign{\hfil$\textstyle##$\hfil\cr
<\cr\sim\cr}}}
{\vcenter{\offinterlineskip\halign{\hfil$\scriptstyle##$\hfil\cr
<\cr\sim\cr}}}
{\vcenter{\offinterlineskip\halign{\hfil$\scriptscriptstyle##$\hfil\cr
<\cr\sim\cr}}}}}
\def\ga{\mathrel{\mathchoice {\vcenter{\offinterlineskip\halign{\hfil
$\displaystyle##$\hfil\cr>\cr\sim\cr}}}
{\vcenter{\offinterlineskip\halign{\hfil$\textstyle##$\hfil\cr
>\cr\sim\cr}}}
{\vcenter{\offinterlineskip\halign{\hfil$\scriptstyle##$\hfil\cr
>\cr\sim\cr}}}
{\vcenter{\offinterlineskip\halign{\hfil$\scriptscriptstyle##$\hfil\cr
>\cr\sim\cr}}}}}

\def\ref[#1]{\noindent \hangindent=1.5em {\bf #1}}
\def\AA#1,#2.{{\it A\&A}, {\bf #1}, #2.}
\def\AAL#1,#2.{{\it A\&A}, {\bf #1}, L#2.}
\def\AAS#1,#2.{{\it A\&AS}, {\bf #1}, #2.}
\def\AJ#1,#2.{{\it AJ}, {\bf #1}, #2.}
\def\APJ#1,#2.{{\it ApJ}, {\bf #1}, #2.}
\def\APJL#1,#2.{{\it ApJ}, {\bf #1}, L#2.}
\def\APJS#1,#2.{{\it ApJS}, {\bf #1}, #2.}
\def\ARAA#1,#2.{{\it ARA\&A}, {\bf #1}, #2.}
\def\BAAS#1,#2.{{\it BAAS}, {\bf #1}, #2.}
\def\MN#1,#2.{{\it MNRAS}, {\bf #1}, #2.}
\def\NAT#1,#2.{{\it Nat}, {\bf #1}, #2.}
\def\SCI#1,#2.{{\it Sci}, {\bf #1}, #2.}
\def\AASIN#1,#2.{{\it Acta Astron. Sin.}, {\bf #1}, #2.}
\def\CAA,#1,#2.{{\it Chin. Astron. Astrophys.}, {\bf #1}, #2.}

\def\PNAS#1,#2.{{\it Proc. Nat. Acad. Sci. USA}, {\bf #1}, #2.}
\def\PASJ#1,#2.{{\it PASJ}, {\bf #1}, #2.}
\def\PASP#1,#2.{{\it PASP}, {\bf #1}, #2.}
\def\IAUEX,#1.{{in: {\it IAU Symposium 97, Extragalactic Radio
    Sources}, ed.\ D.~S. Heeschen and C.~M. Wade (Dordrecht: Reidel),
    p.~#1.}}
\def\IAUVLB,#1.{{in: {\it IAU Symposium 110, VLBI and Compact Radio 
    Sources}, ed.\ R.~Fanti, K.~Kellermann, and G.~Setti (Dordrecht:
    Reidel), p.~#1.}}
\def\IAUQSO,#1.{{in: {\it IAU Symposium 119, Quasars},
    ed.\ G.~Swarup and V.~K. Kapahi (Dordrecht: Reidel), p.~#1.}}
\def\IAUOBS,#1.{{in: {\it IAU Symposium 121, Observational Evidence
    of Activity in Galaxies}, ed.\ E.~Khachikian, G.~Melnick, and
    K.~Fricke (Dordrecht: Reidel), p.~#1.}}
\def\IAUIMP,#1.{{in: {\it IAU Symposium 129, The Impact of VLBI on
    Astrophysics and Geophysics},
    ed. M. J. Reid and J. M. Moran (Dordrecht: Kluwer), p.~#1.}}
\def\IAUACT,#1.{{in: {\it IAU Symposium 134, Active Galactic Nuclei},
    ed. D. E. Osterbrock and J. S. Miller (Dordrecht: Kluwer), p.~#1.}}
\def\IAUGAL,#1.{{in: {\it IAU Symposium 136, The Center of the Galaxy},
    ed. M. Morris (Dordrecht: Kluwer), p.~#1.}}
\def\slm#1.{{in: {\it Superluminal Radio Sources}, ed.\ J.~A. Zensus
    and T.~J. Pearson (Cambridge University Press), p.~#1.}}
\def\pcjet#1.{{in: {\it Parsec-scale radio jets}, ed.\ J.~A. Zensus
    and T.~J. Pearson (Cambridge University Press), p.~#1.}}
\def\nobeya#1.{{in: {\it Frontiers of VLBI}, Frontiers Science Series
    No.\ 1, ed.\ H. Hira\-ba\-yashi, M. Inoue and K. Kobayashi (Universal 
    Academy Press, Tokyo), p.~#1.}}
\def\turku#1.{{in: {\it Variability of Blazars}, ed.\ E. Valtaoja and
    M. Valtonen (Cambridge University Press), p.~#1.}}
\def\heidel#1.{{in: {\it Physics of Active Galactic Nuclei}, ed.\ W.J.
    Duschl and S.J. Wagner (Springer, Heidelberg), p.~#1.}}
\def\ringjet#1.{{in: {\it Jets in Extragalactic Radio Sources}, ed.\ 
H.-J. R\"oser and K. Meisenheimer (Springer, Heidelberg), p.~#1.}}
\def\sara#1.{{in: {\it Sub Arcsecond Radio Astronomy}, ed.\ R.J. Davis
    and R.S. Booth (Cambridge University Press), p.~#1.}}
\def\ringsgr#1.{{in: {\it The Nuclei of Normal Galaxies: Lessons from the Galactic Center}, 
    ed.\ R. Genzel (Kluwer, Dordrecht), p.~#1.}}
\def\iaugenf#1.{{in: {\it IAU\,159:  Multi-Wavelength Continuum Emission of AGN}, 
ed.\ T.J.-L. Courvoisier and A. Blecha, (Kluwer, Dordrecht), p.~#1.}}
\def\geoastro#1.{{in: {\it Proceedings of the 9\,th working meeting on European VLBI
    for Geodesy and Astrometry}, ed.\ J. Campbell and A. Nothnagel, (Bonn), p.~#1.}}
\def\cers#1.{{in: {\it Compact Extragalactic Radio Sources}, ed. J.A. Zensus
    and K.I. Kellermann (NRAO, Socorro), p.~#1.}}
\def\japanvlbi#1.{{in: {\it VLBI Technology, Progress and Future Observational
    Possibilities}, ed.\ T. Sasao, S. Manabe, O. Kameya, and M. Inoue (Terra 
    Scientific Publishing Company, Tokyo), p.~#1.}}
\def\torun#1.{{in: {\it Proceedings of the 2$^{nd}$ EVN/JIVE Symposium}, ed.\ A.J. Kus,
    R.T. Schilizzi, K.M. Borkowski and L.I. Gurvits (Torun Radio Astronomy Observatory,
    Poland), p.~#1.}}

\def\cygwork#1.{{in: {\it Cygnus\,A: Study of a Radio Galaxy}, ed.\  C.L. Carilli
    and D.E. Harris, (Cambridge University Press, Cambridge), p.~#1.}}
\makeatother

\begin{document}
\begin{titlepage}
\bc
{\huge \bf
Millimeter-VLBI with a Large Millimeter-Array: Future Possibilities\\
}

\vspace{3cm}
{\large Thomas P. Krichbaum\\
}

\vspace{3cm}
{\small
Max-Planck-Institut f\"ur Radioastronomie, Auf dem H\"ugel 69, D-53121 Bonn, Germany\\
}
\ec

\vfill
\n
appeared in: `Science with Large Millimeter Arrays', ESO Astrophysical Symposia,
ed.\ P.A. Shaver, Springer (Berlin Heidelberg), 1996, p. 95-102 (ISBN 3-540-61582-2).
\end{titlepage}

\pagenumbering{arabic}
\title{Millimeter-VLBI with a Large Millimeter-Array: Future Possibilities}

%\author{Thomas P. Krichbaum\inst{1}}
\author{Thomas P. Krichbaum}

\institute{Max-Planck-Institut f\"ur Radioastronomie, Auf dem H\"ugel 69,
D-53121 Bonn, Germany}

\maketitle

\begin{abstract}
We discuss possibilities and improvements which could be obtained,
if a phased  array with a large number (N$=50-100$) of sub-millimeter antennas -- like the planned 
large southern array (LSA) -- is used for radio-interferometry
with very long baselines (VLBI) at millimeter wavelengths. We find that the addition
of such an instrument will push the detection limit and the imaging capabilities of a global
mm-VLB-interferometer by 1-2 orders of magnitude. 
\end{abstract}

\section{Introduction}

The \underline{\bf V}ery \underline{\bf L}ong \underline{\bf B}aseline 
\underline{\bf I}nterferometry (VLBI-) technique 
(eg.\  Thompson, Moran \& Swenson, 1986; Zensus, Diamond \& Napier, 1995)
is used to study compact
radio sources. The classes of objects which can be studied at centimeter wavelengths
range from the ultra-luminous nuclei of radio-loud quasars, blazars and BL\,Lac-objects and
their emanating jets, to the moderately luminous radio-galaxies, Seyfert- and starburst galaxies. 
With milli-Jansky sensitivity at centimeter-wavelengths, also the mapping  
of peculiar binary stars (eg. Cyg\,\-X-3, SS\,433), radio-stars and young supernova remnants 
(eg. SN\,1993J) now is possible.  Spectroscopic VLBI-observations allow
to investigate compact maser emitting regions of various kinds (eg. OH--, CH$_3$OH, H$_2$O, SiO) in galaxies, 
star birth regions and stellar envelopes. Polarization-VLBI observations show the
polarized structure of compact radio sources on milli-arcsecond scales. The phase-referencing
technique allows to determine source positions with highest possible accuracy. In geodesy,
VLBI is used to derive earth-rotation parameters and to measure the motion of tectonic plates.

At millimeter wavelengths ($\lambda \leq 0.7$ mm, $\nu \geq 43$ GHz)  present
VLBI-observations allow imaging with an angular resolution of up to
40 $\mu$as (1 $\mu$as $\cor 10^{-6}$ arcsec) at 3\,mm wavelength (eg. B\aa \aa th \etal, 1992). 
This gives the opportunity to study physical processes in regions of micro-arcsecond size,
corresponding to a few up to a few hundred lightdays in a source at cosmological distance $z \geq 0.01$.

To date mm-VLBI is done mostly with telescopes designed for observations at
longer wavelengths. This and the limitations set by not yet fully optimized receivers
($T_{\rm sys}$ in the range of $200 - 1000$ K at 86\,GHz), still limit the detection
sensitivity to about $0.2-0.5$ Jy in recent 3\,mm-VLBI observations (eg. Standke \etal, 1994,
Schalinski \etal, 1994). The addition of sub-millimeter telescopes (eg. the IRAM
30\,m telescope on Pico Veleta, Spain) has helped considerably to push the detection 
thresholds below the 1\,Jy level (eg. Krichbaum \etal, 1993 \& 1994),
but antenna apertures of typically $10^4$ m$^2$ are needed to reach the milli-Jansky (mJy-) level,
which now is accessible at the longer cm-wavelengths without major efforts.

\section{Why millimeter-VLBI ?}

\paragraph{Opacity:}

Most compact extragalactic radio sources are partially 
self-absorbed at cm-wave\-lengths. For a given brightness
temperature $T_{\rm B}$ and flux density $S$ of the radio source, the size $\theta$ of the
emitting region only depends on the observing wavelength $\lambda$:
\begin{equation}
\theta \propto \lambda \cdot \sqrt{\frac{S}{T_{\rm B}}}
\end{equation}
Millimeter-VLBI observations  therefore provide twofold advantage: high angular resolution and
imaging of small scale regions, which are self-absorbed at longer wavelengths and
which cannot be studied directly by other methods.

Recently the frequency band accessible for mm-VLBI was extended to $\lambda=1.4$ mm
($\nu = 215$ GHz) (Greve \etal, 1995).
This and earlier studies with US-american instruments at 223 GHz (Padin \etal, 1990)
and at 86 GHz (Rogers \etal, 1984) indicate that the sources are sufficiently bright and compact 
and that the limitations set by the atmosphere (coherence) still allow VLBI observations 
up to frequencies of at least 300 GHz.

Assuming brightness temperatures below the inverse Compton-limit
($T_{\rm B} \leq 10^{12}$ K) equation (1) yields a minimum size for a source of flux density $S$ (in [Jy]) of
$\theta > (1.22 \cdot S \cdot \nu^{-2})^{1/2} = 4 \cdot \sqrt{S}~ \mu$as at $\nu=300$ GHz.
The rapid flux density variations of the `intraday variable' (IDV-) compact radio sources suggest
that the variable component in this  class of objects is even more compact (eg. Wagner \& Witzel, 1995).
This has the important consequence that at millimeter wavelengths objects with the highest brightness 
temperatures will still appear unresolved on interferometer baseline lengths of $<20\,000$ km. Less compact 
radio sources, however, 
with more typical brightness temperatures in the range of $T_{\rm B}= 10^{9-11}$ K will be partially resolved
by the interferometer beam. It is therefore possible 
to study their underlying structure with unprecedented accuracy.
\begin{table}[t]
\bc
\begin{tabular}{c|c|c|c|c|c} 
$\lambda$ &$\nu$  &A$_{\rm 8000\,km}$ &  $z=1$  & $z=0.01$ &  $r=10$\,kpc \\
$[$mm$]$  &$[$GHz$]$& $[\mu$as$]$ &$[$mpc$]$&$[$mpc$]$&$[\mu$pc$]$ \\ \hline
  6.9     &  43   &   90          &   380   &    12.9  &    4.4 \\
  3.5     &  86   &   45          &   199   &     6.4  &    2.2 \\
  1.4     & 215   &   18          &    77   &     2.6  &    0.9 \\
  0.9     & 350   &   11          &    47   &     1.6  &    0.5 \\  \hline
\multicolumn{3}{l|}{}             &         &          &         \\
\multicolumn{3}{l}{linear resolution in [cm]:}& \multicolumn{1}{|l}{$10^{17-18}$} & \multicolumn{1}{|l}{$10^{15-16}$} & \multicolumn{1}{|l}{$10^{12-13}$} \\
\multicolumn{3}{l|}{}             &         &          &         \\
\multicolumn{3}{r}{in [R$_{S}$]:}& 
\multicolumn{1}{|l}{$10^{3-4}$R$_S^9$}& 
\multicolumn{1}{|l}{$10^{2-3}$R$_S^8$} & 
\multicolumn{1}{|l}{$10^{1-2}$R$_S^6$} \\
\end{tabular}
\ec
\caption{Angular and spatial resolution attainable with ground based mm-VLBI. 
The table gives the typical observing beam size  $A$
(equation (2)) for a 8000\,km interferometer baseline (col.\ 3) depending on 
wavelength (col.\ 1) and frequency (col.\ 2). In columns 4--6 the
corresponding spatial scales  are given for some typical distances of the objects
(using the redshift as distance indicator and \hub). In the lower part of the table the
spatial scale is shown in units of [cm] and Schwarzschild-radii [R$_S$].
With R$_S^9$ being the Schwarzschild-Radius for a $10^9$M$_{\solar}$mass black
hole, a scale of 1 mpc ($= 3.1 \times 10^{15}$ cm) corresponds to $\simeq 10$R$_S^9$.
}
\vspace{-0.5cm}
\end{table}

\paragraph{Angular and spatial resolution:}

The angular resolution $A$ of a radio interferometer is proportional to the observing wavelength
$\lambda$ and inversely proportional to the relative separation or baseline $b$ between two antennas:
\begin{equation}
A \propto \frac{\lambda}{b}~~,{\rm ~or~in~convenient~units:}~~
A_{\rm [mas]} = 3.1 \cdot 10^4 \cdot \frac{1}{\nu_{\rm [GHz]}
\cdot b_{\rm [km]}}
\end{equation}
Since $b$ is limited by the earth's diameter, the only way to achieve higher angular resolution
is to go to shorter wavelengths ($\lambda \rightarrow$ millimeter wavelengths) or to place one or
more VLBI antennas in space (or on the moon). Efforts in both directions are underway:

Recently fringes with significant signal-to-noise ratios of up to $10$ have been
detected for the quasars 3C273, 3C279 and 2145+067
at $\lambda = 1.4$ mm ($\nu = 215$ GHz) on the $1150$ km baseline between the $30$ m
millimeter radio-telescope at Pico Veleta (near Granada, Spain) and a single antenna
of the millimeter radio-interferometer at Plateau de Bure (near Grenoble, France)
(Greve \etal, 1995). Further VLBI test-experiments at frequencies $\nu \geq 150$ GHz are planned.

In table 1 the angular and spatial resolution, which now or in the near future could be achieved 
by ground based mm-VLBI is summarized. Depending on the distance to the source, spatial scales of order
of $10^2-10^4$ Schwarzschild-radii of a $10^9$M$_{\solar}$mass black hole can be imaged
in extragalactic sources. For galactic sources (eg. the Galactic Center Source Sgr\,A*) regions
of 10-100 Schwarzschild-radii (assuming a $10^6$M$_{\solar}$mass black hole) can be mapped.
This is close to the sizes of the expected central accretion disc !

Space VLBI at cm-wavelengths ($\lambda \geq 1.3$ cm) with an orbiting VLBI antenna is planned 
for the very near future (the Japanese satellite `VSOP' will be launched in fall 1996). 
If successful, this will trigger further technical improvements and future missions. With an open mind for 
such future developments, it therefore is not unreasonable to assume that on timescales on which also 
a large sub-millimeter array like the LSA would be completed, space-VLBI observations might be quite common. 
One therefore should regard mutual observations between earth- \underline{and} 
space-based telescopes, even at millimeter wavelengths. Millimeter space-VLBI  would yield, for example, 
an angular resolution of $\sim 4~ \mu$as, if interferometric 
observations on a $20\,000$ km baseline are performed at
$\lambda = 1$ mm ($\nu = 300$ GHz). Even if space-VLBI at the shortest millimeter wavelengths were not 
feasible by the time the LSA is operating, ground based mm-VLBI observations could complement 
space-VLBI observations at short cm-wavelengths with their {\it matching} angular resolution.
Thus small scale regions could be imaged with nearly identical resolution at quite
different frequencies (eg. 86\,GHz space-VLBI and 215 GHz ground-VLBI observations yield similar
observing beams).
\begin{table}[t]
\vspace{-0.5cm}
\bc
\begin{tabular}{|l|l|c|c|c|c|} \hline
Station Location            &Abbrev.     &  D    & T$_{\rm sys}$   &$\eta_A$      & G$_{\rm eff}$ \\
                            &            &$[m]$  &   ~$[K]$~       &~$[\%]$~      &~$[K/Jy]$~~\\ \hline
Large Southern Array, type 1&LSA1        &50x16  &     150         &  0.6         & 2.16~    \\
Large Southern Array, type 2&LSA2        &100x11 &     150         &  0.6         & 2.06~    \\
NRAO Millimeter Array       &MMA         & 40x 8 &     150         &  0.6         & 0.43~   \\  
Large Millimeter Telescope, Mexico  
                            &LMT         & 50    &     150         &  0.3         & 0.21~   \\ \hline
Plateau de Bure, France     &Bure        & 5x15  &     150         &  0.3         & 0.086     \\ 
Pico Veleta, Spain          &Pico        & 30    &     150         &  0.3         & 0.077     \\
Nobeyama, Japan             &NRO         & 45    &     150         &  0.1         & 0.058    \\
Nobeyama Millimeter Array, Japan  
                            &NMA         & 6x10  &     150         &  0.3         & 0.047     \\
Owens Valley, California    &Ovro        & 6x10.4&     150         &  0.5         & 0.084     \\
Hat Creek, California       &Bima        &10x6.1 &     150         &  0.5         & 0.050     \\
Southern European Telesc.,  Chile
                            &Sest        & 15    &     150         &  0.35        & 0.022     \\
James Cark Maxwell Telesc., Hawai 
                            &JCMT        & 15    &     150         &  0.6         & 0.038     \\
Kitt Peak, Arizona          &KittP       & 12    &     150         &  0.4         & 0.016     \\
Heinrich Hertz Telescope, Arizona
                            &HHT         & 10    &     150         &  0.6         & 0.017     \\ 
Caltech Sub-mm Observatory, Hawai
                            &CSO         & 10.4  &     150         &  0.5         & 0.015     \\ \hline
\end{tabular}
\ec
\caption{A hypothetical future VLBI-array at $\sim 250-300$ GHz. Not all possible participants
are listed. Being very optimistic we assumed 
for each antenna an effective system temperature $T_{\rm sys}=
150$ K (including atmosphere). The effective antenna gain 
G$_{\rm eff}$ in [K/Jy] is shown in column 6, and follows from 
G$_{\rm eff} \simeq 2.845 \cdot 10^{-4} \cdot \eta_A \cdot D^2$, with the telescope diameter D in [m]. 
Already existing mm-antennas which are expected to contribute with reduced gain
are listed in the lower part of the table.
}
%\end{table}
%\clearpage
%\begin{table}[b]
%\vspace{0.5cm}
\vfill
\bc
\begin{tabular}{llr}
LSA & LSA/MMA/LMT                      &  3--10\,mJy \\
LSA & Ovro/Bure/Pico/Nobe/Bima/NMA     & 15--20\,mJy \\
LSA & JCMT/Sest/HHT/KittP/CSO          & 20--35\,mJy \\
MMA & LMT/Ovro/Bure/Pico/Nobe/Bima/NMA & 20--40\,mJy \\
LMT & Bure/Ovro/Pico/Bima/Nobe/NMA     & 40--60\,mJy \\
    & others                           &$\ga$ 60\,mJy \\
\end{tabular}
\ec
\caption{Typical $7\sigma$ VLBI detection thresholds. We assumed 2-bit sampling,
1\,GHz bandwidth, 20\,sec integration time, and a signal-to-noise ratio of the
detection of at least SNR $\geq 7$. The flux density limits were derived using table 2 and 
equation (3).}
\vspace{-0.5cm}
\end{table}

\section{Sensitivity}

The single baseline $1\,\sigma$-detection threshold (in [mJy]) of a VLBI-interferometer consisting
of two antennas with diameters $D_i$ and $D_j$ (in [m]), aperture efficiencies $\eta_i$
and $\eta_j$ observing with receivers of system temperatures $T_{\rm sys}^i$ and 
$T_{\rm sys}^j$ (in [K]) at a bandwidth $\Delta \nu$ (in [MHz]) is:
\begin{equation}
\sigma_{ij} =  2.485 \cdot 10^6 \cdot \frac{1}{C_l} \cdot \frac{1}{D_i \cdot D_j} \cdot 
\sqrt{ \frac{T_{\rm sys}^{~i}}{\eta_i} \cdot \frac{T_{\rm sys}^{~j}}{\eta_j}}
\cdot \frac{1}{\sqrt{\Delta \nu \cdot \Delta t}}
\end{equation}
where $C_l$ is a VLBI efficiency factor combining quantization and correlator losses (eg. $C_l = 0.88$
for 2-bit sampling), and $\Delta t$ is the integration time (in [sec]).
At mm-wavelengths coherence losses in the atmosphere limit the integration time $\Delta t$ to 
a few up to a few tens of a second (note: incoherent averaging of the coherent segments in
the initial fringe search allows detection of fringes somewhat beyond the atmospheric
coherence time (Rogers \etal, 1995)). The limits for
single antenna sizes ($D \leq 100$ m), aperture efficiencies ($\eta \la 0.6$),
and receiver performances ($T_{\rm RX} \ga 50-100$ K for $\nu \geq 50$ GHz) do not allow
a substantial lowering of the detection threshold. Some improvement could be obtained from the 
extension of the observing bandwidth, but a major step could only be done with an increase of
the effective collecting area. 

The MK\,IV and VLBA data acquisition systems will provide data recording with extended
bandwidths of up to $\Delta \nu=512$ MHz in the foreseeable future. But even if observations
with GHz-bandwidth were possible (and if the problems of phasing an array at such a
large bandwidth are solved), the VLBI-detection limit can be pushed by not much more than a factor
of $\sim 3$.

A large collecting area of order of $10^4$ m$^2$ requires an aperture synthesis instrument.
The addition of such a phased array to an existing network of smaller VLBI-telescopes
would lower the VLBI-detection threshold at millimeter wavelengths drastically. As illustrating
example, we list the station performances of a hypothetical future mm-VLBI array in table 2.
We included both suggested configurations for the LSA (50 $\times$ 16\,m antennas (LSA1) or 100 $\times$
11\,m antennas (LSA2)) and assumed also that the proposed NRAO Millimeter Array (MMA) and
the UMASS/INAOE Large Millimeter Telescope (LMT) will be
operating. From equation (2) and with some simplifying assumptions on antenna and 
receiver characteristics, we estimated  the VLBI-detection
limits, which we summarize in table 3. It is obvious that the combination of the LSA with
other large sub-mm telescopes pushes the detection threshold towards the $5-10$ mJy level.
More optimistic assumptions on observing bandwidth 
and receiver performances would even lower these sensitivity limits.

The addition of one very sensitive antenna to a VLB-array consisting mainly of smaller antennas
produces a considerable improvement of the dynamic range of the images (see eg.
C. Walker, 1989): 
\begin{equation}
\frac{1}{\Delta S^2} = \sum_{i,j}^{{\rm all~baselines}} \frac{1}{\sigma_{ij}^2}
\end{equation}
where $\Delta S$ (in [mJy]) is the noise in the map, $\sigma_{ij}$ is taken from equation (3),
and the sum is over all $N(N-1)/2$ baseline combinations of the interferometer array. Replacing
for example one antenna in an array of 5 equal 30\,m antennas by the phased LSA would
cause a reduction in the map noise by a factor of 3 to about $\sigma=1$ mJy. This
and the fact that the single baseline `a-priori' detection threshold between two smaller 
and less sensitive antennas can be reduced, if `fringes' to a third more sensitive antenna are detected
(the method of 'global fringe fitting' uses the closure relations in such antenna triangles, 
eg. Schwab \& Cotton, 1983, Alef \& Porcas, 1986), emphasizes the dramatical improvement of the
data quality, which could be achieved, if the LSA is added to any preexisting VLB-array.

\section{What can be observed ?}

From equation (1) and (2) the lowest detectable brightness temperature can be written as:
\begin{equation}
T_{\rm B} {\rm [K]}  \geq 1.27 \cdot \Delta S_{\rm [mJy]} \cdot b_{\rm [km]}^2 \cdot r^{-2}
\end{equation}
where the ratio $r$ of source size to the interferometer beam 
($r=\theta_{\rm source} / A$) measures the source extent, 
$b$ the maximum baseline length of the interferometer and $\Delta S$ the 
noise level in the map (equation (4)). With a detection limit of $\Delta S \simeq 1$ mJy  and source sizes in
the range $0 \leq r \leq 10$, the lowest detectable brightness temperature at $\nu=300$ GHz is
$T_{\rm B} \geq 10^{2-4}$ K for a 100 km baseline ($A=1$ mas), $T_{\rm B} \geq 10^{4-6}$ K for a 1000 km baseline 
($A=0.1$ mas) and $T_{\rm B} \geq 10^{6-8}$ K for a 10000 km baseline ($A=0.01$ mas). In figure 1 
this result is also displayed graphically.
\begin{figure}[t]
\rotate[r]{\psfig{file=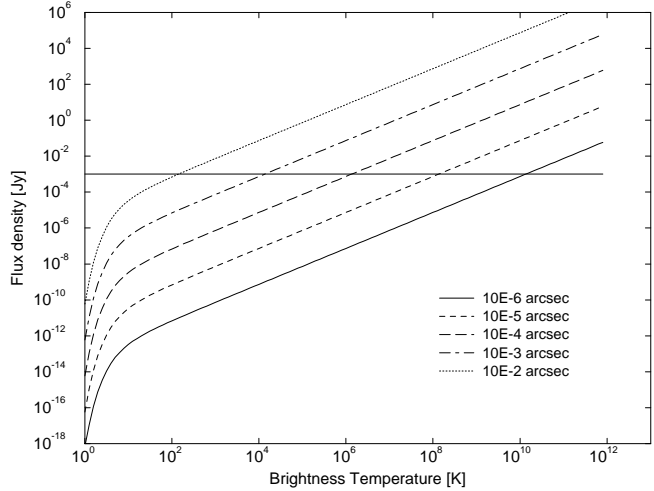,width=8.5cm}}
\caption{Flux density of a Planck-black-body radiator plotted versus brightness temperature.
The curves are for different source sizes ranging from $1 \mu$as ($10^{-6}$ arcsec, lower solid line)
to $10$ mas ($10^{-2}$ arcsec, upper dotted line). The horizontal line gives a hypothetical
VLBI detection threshold of $S_{\rm min} = 1$ mJy. For a given source size, the minimum
detectable brightness temperature can be found at the position were horizontal line and
radiation curve intersect.}
\vspace{-0.5cm}
\end{figure}

Thus it is clear that most compact nonthermal radio sources and the brighter thermal
objects ($T_{\rm B} \ga 10^{5-6}$ K) are accessible for mm-VLBI including the LSA. 
It can be expected that such observations substantially would improve our understanding of the various
forms of nuclear activity in galaxies (eg. starbursts, mergers, central black holes) and quasars 
(accretion, production and propagation of jets, particle acceleration processes). High dynamic
range mm-VLBI-imaging with micro-arcsecond resolution will reveal more insight in
the central light-day size regions of these objects, self-absorbed at longer wavelengths. This should 
result in a much more detailed understanding of the still unsolved problem of energy production 
in active galactic nuclei (AGN).

VLBI imaging of high redshift quasars (z $\geq 3$) and gravitational lenses at mm-wavelengths can help
in answering questions on the metric and structure of our universe. The extension of the angular 
size-distance relation towards smaller source sizes and larger distances, could help
to more accurately determine the cosmological deceleration parameter $q_0$ (eg.\ Kellerman \etal, 
1993, Gurvits \etal, 1994) or even the cosmological constant $\Lambda$ (Krauss \& Schramm, 1993).

With its inverted radio spectrum ($\alpha >0$, \sprop),
the flux density of thermally emitting objects increase from the cm- to the mm-regime. With a planned
maximum antenna spacing of $\la 10$ km and an angular resolution of $\ga 10$ mas, the LSA will
be particularly useful for detailed studies of low brightness temperature objects $T_{\rm B} \la 100$ K.
At the mJy-level, radio stars with brightness temperatures in the range of $T_{\rm B} =10^{3-5}$ K  
have sizes in the milli-arcsecond range (see Altenhoff \etal, 1994 for more details on radio stars). Thus
studies of stars, stellar winds or even stellar surfaces would become possible,
if the LSA and other sensitive antennas (eg. MMA, LMT) were used as a long baseline interferometer
covering only intermediate baseline lengths in the range of typically $10-1000$ km. 
In the radio cm-bands the thermal part of the spectrum of a 
radio source (eg.\ a radio star) is too faint to be observable with VLBI.
In the mm-bands this part may reach flux densities of order of up to a few mJy and sizes of less than
a few or a few ten  milli-arcseconds.
Millimeter-VLBI observations with the LSA and possibly the MMA and other large antennas
will for the first time allow direct imaging of such objects. 

\section{Summary}

The planned large southern array will dramatically improve the quality of scientific
research at millimeter wavelengths. With its large collecting area it could play an
important role in future mm-VLBI observations, pushing the sensitivity to the milli-Jansky
level. In VLBI, such high sensitivity at present is achieved only at the longer
cm-wavelengths. Since thermal radiation becomes more dominant towards shorter wavelengths, 
mm-VLBI with milli-Jansky sensitivity will not only allow imaging of compact non-thermal
radio sources, but also mapping of thermally emitting objects like stars or hot compact
regions in extragalactic objects. With an angular resolution of micro-arcseconds, high sensitivity
mm-VLBI or even space mm-VLBI (using an orbiting mm-antenna) will allow imaging of
regions not directly accessible by any other method. It is therefore not unreasonable to assume
that this observing method could yield spectacular scientific results, improving our
present knowledge of many astronomical objects in the universe.\\

\n
{\it Acknowledgements: VLBI observations at millimeter wavelengths are a joint effort of numerous
people, impossible to list here. We wish to express thanks to all of them. The following 
observatories were involved in recent 86 GHz VLBI campaigns:
MPIfR (100 m Telescope),
IRAM (Pico Veleta, Plateau de Bure),
NRAO (Kitt Peak),
Onsala,
Sest,
Haystack,
Quabbin,
the Owens Valley Caltech interferometer (OVRO),
and the
Berkeley Hat Creek interferometer (BIMA).
The author appreciates financial support from the German BMFT-Verbundforschung.
}

%
% ---- Bibliography ----
%

\end{document}